\documentclass[%
 aip,
 amsmath,amssymb,
 reprint,%
]{revtex4-1}

\usepackage{graphicx}
\usepackage{dcolumn}
\usepackage{bm}

\usepackage[T1]{fontenc}
\usepackage{mathptmx}
\usepackage{amsmath}
\usepackage{etoolbox}
\usepackage[english]{babel}
\usepackage{tikz}
\usetikzlibrary{shapes.geometric, arrows}

\makeatletter
\def\@email#1#2{%
 \endgroup
 \patchcmd{\titleblock@produce}
  {\frontmatter@RRAPformat}
  {\frontmatter@RRAPformat{\produce@RRAP{*#1\href{mailto:#2}{#2}}}\frontmatter@RRAPformat}
  {}{}
}%
\makeatother
\begin{document}

\preprint{AIP/123-QED}

\title{Pressure-Induced Negative-Positive Magnetoresistance Crossover Near Metal-Insulator Transition in $La_{0.8}Ag_{0.1}MnO_{3}$}
\author{Adler G. Gamzatov}
\affiliation{%
Amirkhanov Institute of Physics of DFRC of RAS, Makhachkala, 367003, Russia
}%
\affiliation{%
National University of Science and Technology “MISIS”, Moscow, 119049, Russia\\
}%

\author{Temirlan R. Arslanov}
 
\affiliation{%
Amirkhanov Institute of Physics of DFRC of RAS, Makhachkala, 367003, Russia
}%

\author{Andrey R. Kaul}
 
\affiliation{%
Department of Chemistry, Lomonosov Moscow State University, Moscow, 119991, Russia
}%

\author{Zaur Z. Alisultanov}%
 \email{zaur0102@gmail.com}
\affiliation{%
Abrikosov Center for Theoretical Physics, MIPT, Dolgoprudnyi, Moscow Region, 141701, Russia
}%

\affiliation{%
Amirkhanov Institute of Physics of DFRC of RAS, Makhachkala, 367003, Russia
}%


\begin{abstract}

We investigated the effect of high pressure on the field dependences of magnetoresistance (MR) in La$_{0.8}$Ag$_{0.1}$MnO$_{3}$ near the metal-insulator transition temperature. Our results showed that an increase in pressure results in a decrease in the magnitude of negative MR. At pressures $P\geqslant5.6$ GPa and magnetic fields up to 4 kOe, we observed a positive MR. However, with a further increase in magnetic field ($>4$ kOe), the MR again became negative. Therefore, we discovered a "negative-positive" MR crossover induced by high pressure near the transition temperature. We supported our experimental findings with a qualitative theoretical interpretation  using the electron-hole model of MR. This theory explains observed the MR sign change.

\end{abstract}

\maketitle

The influence of magnetic field (H) and pressure (P) on the magnetic, transport and thermophysical properties of manganites are often similar~\cite{Dagotto,Markovich}. For example, pressure affects both the volume of the elementary cell and the local structure, significantly changing the magnetic and transport properties of the manganites, that, like a magnetic field, leads to increasing of $T_{MI}$ (metal-insulator phase transition temperature) and $T_c$~\cite{Markovich}. While, it should be noted that the mechanisms of pressure and magnetic field effects on the charge carriers transport are significantly different. The magnetic field, by aligning the spins between adjacent Mn ions, increases the bandwidth (W) of free electrons and, consequently, increases mobility through the double exchange mechanism. At the same time, the effects of pressure on the Curie temperature and magnetotransport (at a fixed concentration x) differ significantly depending on whether it is external or internal chemical pressure (due to the replacement of A-site cations in ABO$_3$ perovskites by small-radius ions)~\cite{Hwang1,Hwang2}. Although in both cases the elementary cell size decreases, internal chemical pressure causes a decrease in the $T_{MI}$ and $T_c$, while the external pressure leads to an increase in both $T_{MI}$ and $T_c$~\cite{Hwang2}. This behavior is related to the fact that chemical pressure reduces the W due to the decrease of valence angle of Mn-O-Mn, while external pressure increases this one~\cite{Laukhin}.

Static and/or dynamic distortions of MnO$_6$ octahedra caused by the Jahn-Teller effect on Mn$^{3+}$ ions play a significant role in the formation of magnetotransport properties of manganites near the phase transition. For example, in the paramagnetic state, Jahn-Teller distortions lead to the self-localization of carriers in the form of polarons, which in turn leads to a dielectric behavior of the resistivity~\cite{Jaime}. An even more complex picture will be observed in ceramic manganites, where intergranular boundaries are very sensitive to external influences (pressure, magnetic field). Thus, studies of magnetotransport properties under high pressure can provide additional information about the balance between structure, magnetism and electron mobility. Simultaneous influence of pressure and magnetic field on magnetotransport properties can lead to interesting effects and provide additional information about transport mechanisms in manganites.

Studies of the influence of external pressure on the magnetotransport in manganites are mainly limited to a pressure of about 2 GPa~\cite{Dagotto,Markovich}. In this paper, we study the hydrostatic pressure effect up to 8 GPa on the magnetoresistance (MR) of La$_{0.8}$Ag$_{0.1}$MnO$_{3}$ manganite. The magnetotransport properties of the La$_{(1-x)}$Ag$_{x}$MnO$_{3}$ system were studied in detail in our early works~\cite{Kamilov2007,Gamzatov2013,Gamzatov2013+}. The temperature behavior of the resistivity has a characteristic form for manganites with a metal-insulator transition at $T_{MI}=296$ K. In the high-temperature paramagnetic phase, the behavior of $\rho(T)$ is described based on the concept of small-radius polarons and is described by the thermal activation law of the following form: $\rho_{PM}=DT\exp(E_{p}/k_{B}T)$, where $D=2k_{B}/(3ne^{2}a\nu)$ (here $n$ is the charge carrier concentration, $a$ is the length of the polaron hop, which roughly coincides with the lattice constant, $\nu$ is optical phonon speed), $E_{p}$ is the activation energy of the polaron jump, which for the La$_{(1-x)}$Ag$_{x}$MnO$_{3}$ system ranges from 100-144 meV~\cite{Kamilov2007}. Studies of the influence of pressure on the resistivity showed that an anomaly is observed in the baric dependence of the resistivity at a temperature of 296 K in the form of a break in the linear behavior of the logarithm of the baric dependence of the resistance, corresponding to a pressure of 3.85 GPa~\cite{Gamzatov2022}.

This Letter reports on an experimental investigation of MR in La$_{0.8}$Ag$_{0.1}$MnO$_{3}$ at different external pressures near the metal-insulator transition. Surprisingly, we observed a reversal of the MR sign as pressure increased, but this effect was reversed as the magnetic field increased. We propose a simple resistive model with two current carrier components - the electron-hole resistive model - to explain this phenomenon.

The synthesis process of La$_{0.8}$Ag$_{0.1}$MnO$_{3}$ ceramic samples involved the chemical homogenization method using aqueous solutions of La(NO$_{3}$)$_{3}$, AgNO$_{3}$, and Mn(NO$_{3}$)$_{2}$ nitrates. Initially, the nitrate solutions were mixed in the required ratio, and then ashless filters were impregnated with the resulting solution. The impregnated filters were dried at 120°C and burned to obtain the residue. The residue was further burned at a temperature of 600°C for 30 minutes, followed by pressing into tablets and sintering at 1200°C in an oxygen flow at 1 atm for 30 hours.

Compounds of the La$_{(1-x)}$Ag$_{y}$MnO$_{3}$ ($y\leq x$) type differ from other systems, in which divalent metal ions play the role of doping elements. These compounds contain, in addition to $y$ two-charged acceptors (Ag$^{+}$ ions), also ($x-y$) three-charged acceptors (La$^{3+}$ vacancies), which also form a high-conductivity ferromagnetic state, inducing the Mn$^{3+}$ Mn$^{4+}$ transition. Silver appears in the perovskite structure, substituting vacancies in the A sublattice (i.e., in the lanthanum sublattice). Thus, compositions with $x=y$ are compounds with a completely filled A sublattice, in which lanthanum and silver ions are statistically distributed over A positions. Intermediate compositions ($y<x$) are compounds with partly filled A sublattice, in which the remaining positions are vacancies. Such silver-deficient compounds can be obtained with improved transport characteristics since they permit high fritting temperatures without precipitation of metallic silver in view of their high thermodynamic stability as compared to compounds with a higher silver content. In all probability, it is possible to select the optimal ratio of vacancies to implanted Ag$^{+}$ ions, which would ensure the best functional parameters of the resultant ceramic. Direct-current (DC) resistivity and MR measurements were performed using a six-contact method in a Toroid type high pressure apparatus under hydrostatic pressure at room temperature~\cite{Khvostantsev}. 

\begin{figure}
\includegraphics[width=\columnwidth]{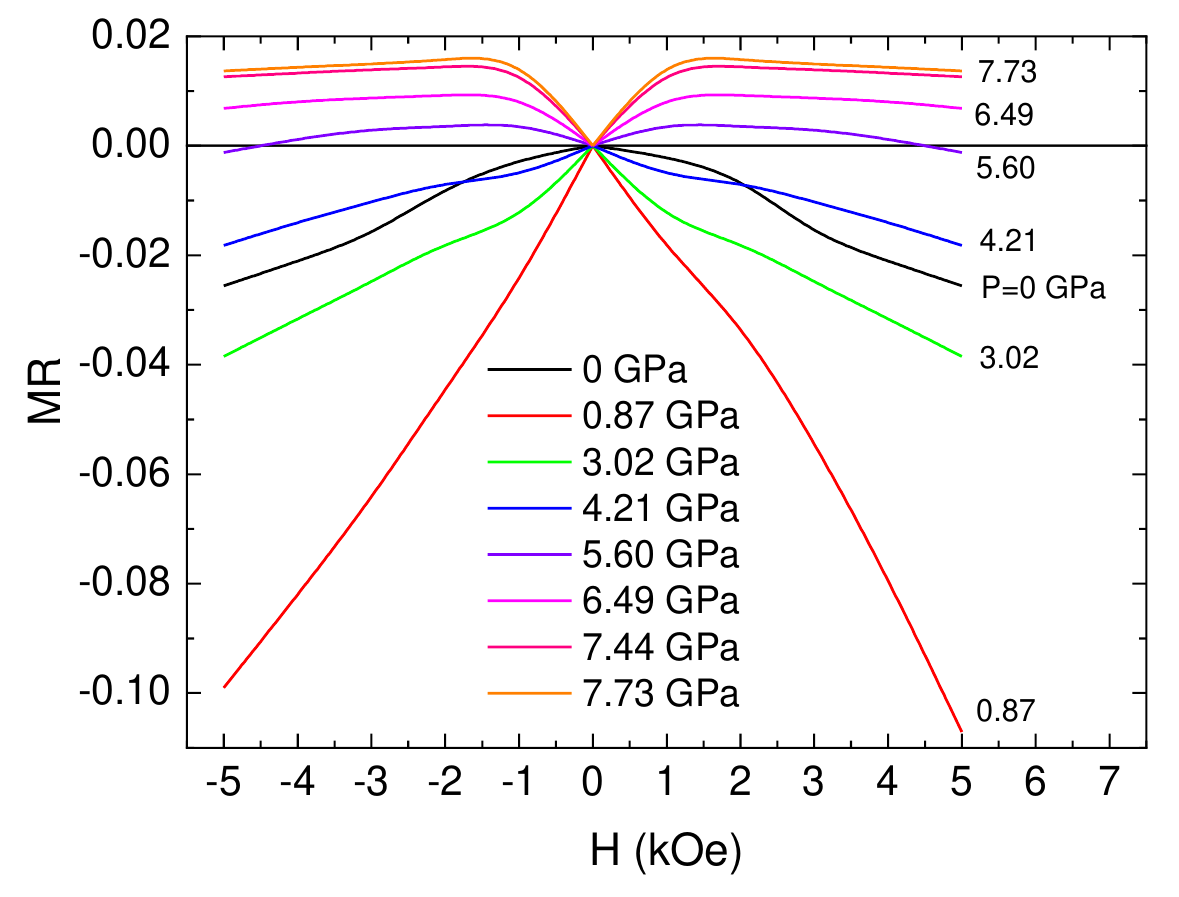}

\includegraphics[width=\columnwidth]{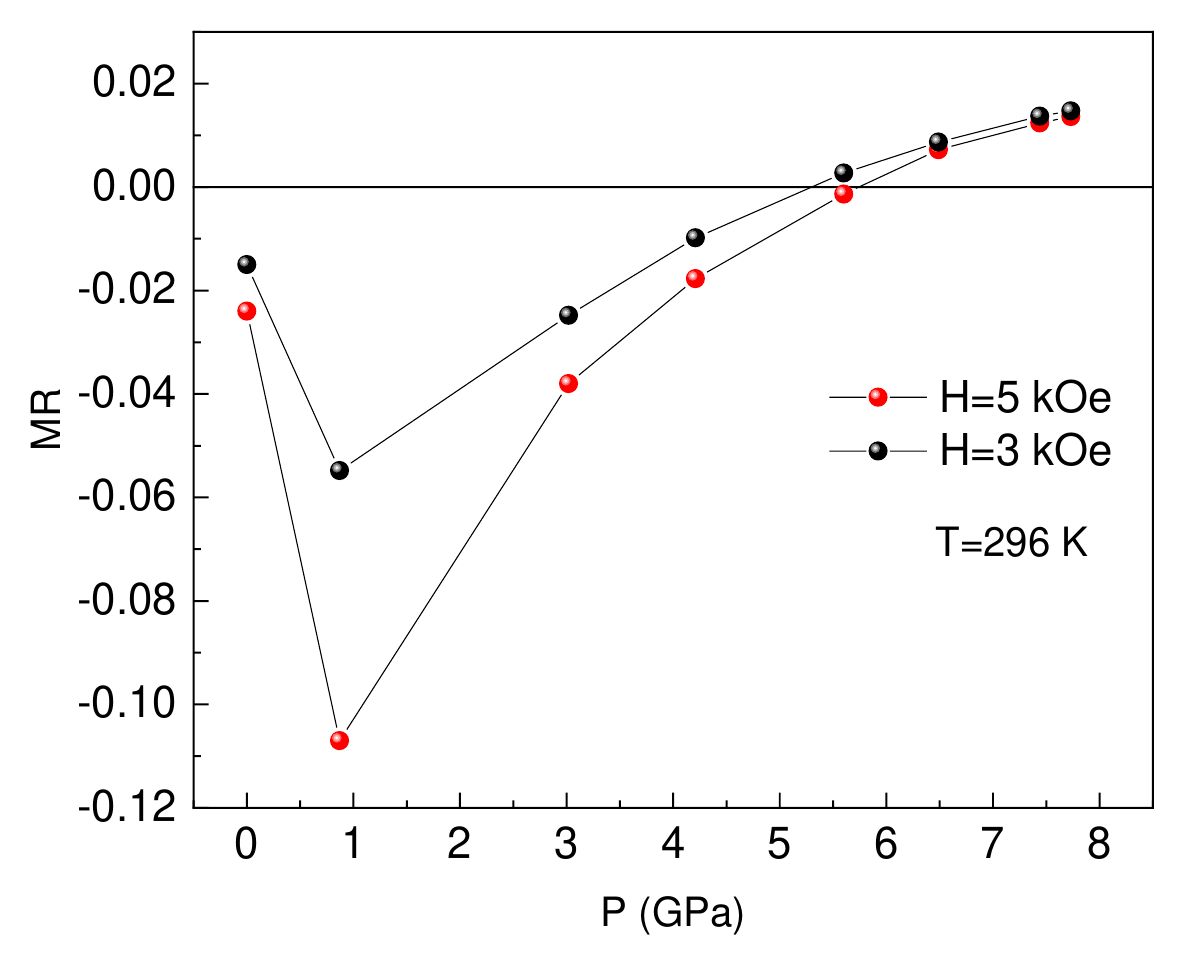}
\caption{a) Magnetoresistance vs. magnetic field for $\text{La}_{0.8}\text{Ag}_{0.1}\text{MnO}_{3}$ at different pressures at T=296 K. b) Magnetoresistance vs. pressure for $H=3$ and 5 kOe.}
\end{figure}

Figure 1(a,b) shows the results of studying the dependence of the magnetoresistivity (MR) on the magnetic field (up to 5 kOe) at different external pressures at T=296 K. The results indicate that an increase in pressure from 0 to 0.87 GPa leads to a fourfold increase in the MR in a 5 kOe field. However, further pressure increase to 3.02 GPa results in the opposite effect, with the MR declining and remaining negative even at 4.21 GPa. At P=5.6 GPa, the MR becomes positive in fields up to 4.5 kOe, indicating a "negative-positive" crossover in the pressure behavior of the MR near T$_C$ (refer to Fig. 1b). This phenomenon is challenging to explain based on the double exchange mechanism and the existence of only one ferromagnetic metallic phase (FMM). Previous studies (refer to~\cite{Markovich2002prb,Markovich2002jap}) have also documented positive MR in manganites, which cannot be explained by assuming the presence of an AFM + FM two-phase magnetic state or the existence of an FMI phase. The presence of a magnetic two-phase state in these manganites was demonstrated in~\cite{Koroleva}. Positive MR is typical of a non-magnetic metal and is attributed to the curvature of charge carriers' trajectory due to the Lorentz force. In this case, MC value is usually negligible since the Hall field compensates for the impact of the Lorentz force on both channels. However, incorporating carriers with an opposite charge in the theory significantly alters the situation. The observed change in MR sign can be explained within the framework of such a two-component theory, as presented below.

Now, we present a straightforward theory of magnetotransport that elucidates the MR crossover phenomenon. While the contemporary theory of magnetotransport in manganites is based on polaron concepts and accurately accounts for temperature dependencies~\cite{Markovich}, our focus lies on the pressure- and magnetic field-induced change in MR sign. For this reason, we employ the language of electrons and holes, as incorporating polaron effects would only complicate matters without explaining the observed crossover. The crucial factor enabling an explanation of the MR sign change is the inclusion of carriers with opposite charge signs (holes) in the theory. Naturally, this electron-hole theory can be rephrased in terms of electronic and hole polarons, enabling the description of temperature dependencies of thermodynamic and transport quantities that are beyond the scope of this article.

The MR is typically composed of two components. The first component is due to the ordering of spins of the ions, resulting in a decrease in resistance, and is typically negative. This component saturates as the magnetic field exceeds a critical value. The second component is the Lorentzian MR, which arises from the Lorentz force acting on current carriers' trajectories and is generally positive. The behavior of the Lorentz MR can vary with changes in magnetic field depending on the material. While it is usually small in metals, it can increase significantly in some cases. The MR can either saturate or diverge as a function of the magnetic field, depending on the Fermi surface topology and the carriers type~\cite{Kittel}.

The conductivity of a magnetic metal consists of two independent channels for spin up and spin down carriers (the scattering probability with spin flip is low in metals)
\begin{equation}
\sigma=\sigma_{\uparrow}+\sigma_{\downarrow}
\label{eq:1}
\end{equation}
According to Mott~\cite{Mott36,Mott64}, valence $sp$-electrons carry most of the electric current due to their low effective mass and high mobility, while $d$-bands play a role in the final scattering states of $sp$ electrons. In ferromagnets, the exchange-split of $d$-bands result in different densities of states for spin-up and spin-down electrons at the Fermi energy, leading to different scattering rates for the two channels. Although this model is simplistic due to strong hybridization between $sp$- and $d$-states, it is useful for understanding spin-dependent conductivity in transition metals and their alloys~\cite{Tsymbal}. 

The spin dependence of conductivity can be understood from the simple Drude theory
\begin{equation}
\sigma_i=\frac{e^{2}}{\pi\hbar}\frac{k_{F}^{2}}{6\pi}\lambda_i
\label{eq:2}
\end{equation}
where $i=\uparrow,\downarrow$, $\lambda_i=\upsilon_{F}\tau_i$ is the mean free path and relaxation time $\tau_i$ can be evaluated by Fermi's golden rule
\begin{equation}
\tau^{-1}_i=\frac{2\pi}{\hbar}\left\langle V_{scat}^{2}\right\rangle n_i\left(E_{F}\right)
\label{eq:3}
\end{equation}
Here $\left\langle V_{scat}^{2}\right\rangle$ is the mean value of the scattering potential, and $n_{\uparrow,\downarrow}\left(E_{F}\right)$ is the density of electronic states (DoS) at the Fermi energy $E_{F}$ for the corresponding spin projection. The spin dependent DoS is crucial in understanding why scattering in ferromagnetic metals is spin-dependent. Other quantities in equations (\ref{eq:2}-\ref{eq:3}) do not exhibit significant spin dependence. A change in the DoS, caused by external perturbation (pressure, in our case) leads to a change in the relaxation time. Our work assumes that external pressure significantly increases the relaxation time, which is supported by other experimental studies (see, e.g.~\cite{pressure1,pressure2,pressure3,pressure4} and also our previous work~\cite{Gamzatov2022}). This assumption, along with the presence of two types of carriers (electrons and holes), serves as a fundamental basis for our theoretical analysis.
\begin{figure}
\includegraphics[width=\columnwidth]{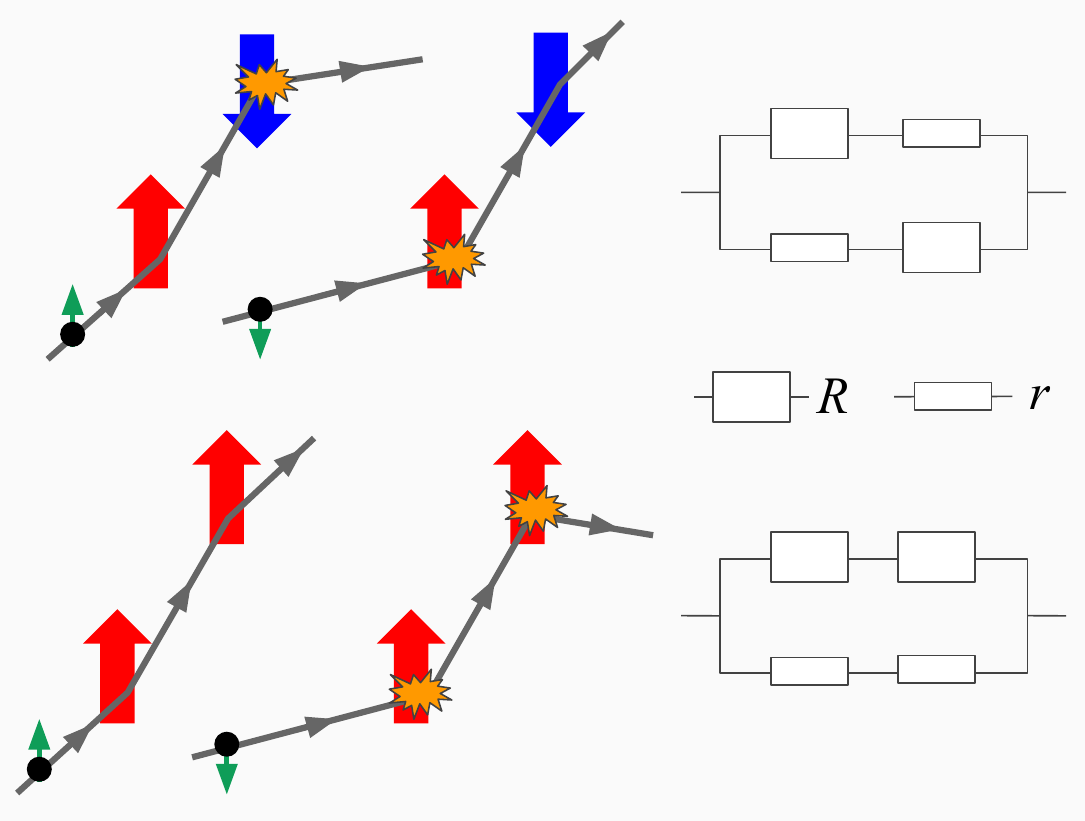}

\caption{The Resistor Model of MR describes the behavior of a system with and without magnetic field. In the upper part of the figure system is not polarized and the scattering for both spin up and spin down channels is equal. The equivalent resistor circuit for this scenario is shown in the top right part. In the lower part of the figure where the system is polarized, the scattering for the channels differs. An equivalent resistor circuit is shown in the bottom right.}
\end{figure}

In this study, we use a resistor model and extend it to the specific system being analyzed. While this model may not provide a precise quantitative explanation of the phenomena being studied, it is still really effective in providing a qualitative understanding. The resistor model states that every magnetic domain, or magnetic moment of the scattering center, is treated as a separate resistor (as shown in Fig. 2). According to this model, scattering with antiparallel spins is denoted as $R$, while scattering with parallel spins is denoted as $r$. Such a simple model gives the following expression for MR
\begin{equation}
MR=-\frac{\left(r-R\right)^{2}}{\left(r+R\right)^{2}}=-\frac{\left(\alpha-1\right)^{2}}{\left(\alpha+1\right)^{2}},
\end{equation}
where $\alpha=r/R\leq1$. We can see that at $\alpha=1$ (identical channels), the MR disappears. Moreover, it can be seen that MR is always negative, as we discussed in the introduction to this chapter. In a real magnetic system, the parameter $\alpha$ varies with the strength of the magnetic field. As the field increases, the spins align more strongly with it, causing $\alpha$ to decrease. When the field reaches saturation value, $\alpha$ no longer changes significantly with further increases in the field. However, the field dependence of $\alpha$ is significant in the low fields.

Let us generalize the four-resistor model (Fig. 2) to the case of an arbitrary number of scattering centers. Let, on average, an electron with spin up encounter N magnetic moments up and M ones down on its way. Situation with $M\neq N$ corresponds to case when magnetic field polarizes the system partially. For an electron with spin down, M and N are reversed. It is clear that $N_{0}=N+M$ is the total number of spin moments. In the absence of magnetic field, $N=M=N_{0}/2$. When the system is polarized, $N=N_{0}/2+\triangle N$ and $M=N_{0}/2-\triangle N$. The value $\triangle N$ is the magnetization in units of individual spin moments and it is a function of the magnetic field. It should be noted that external pressure leads to a rearrangement of the crystal lattice and, consequently, to a change in the exchange integral. As a consequence, this can lead to a partial change of magnetic order. Therefore, in the general case, we write this quantity as $\triangle N\left(B,\triangle J\right)$, where $\triangle J$ is a change in the exchange integral caused by external pressure. In this model, resistance is defined as $Nr+MR$ for a spin up channel and $NR+Mr$ for a spin down channel. Then we obtain the following expressions for the total resistance
\begin{align}
R_{B}=\frac{\left(Nr+MR\right)\left(NR+Mr\right)}{\left(N+M\right)\left(r+R\right)}\\
R_{B=0}=\frac{N_{0}\left(R+r\right)}{4}
\end{align}
The value $\triangle N\left(B\right)$ can be approximated by the function
\begin{equation}
\triangle N\left(B\right)=\frac{N_{0}}{2}\left(1-e^{-\gamma B}\right)
\end{equation}
where $\gamma$ is some constant that depends on the properties of the material. From this dependence, we can conclude that for $B \gg\gamma^{-1}$, the vast majority of moments are oriented along the field. The value $B_{c}=\gamma^{-1}$ can be called the saturation field. It can be seen from this formula that for small fields $\triangle N\left(B\right)\approx\frac{N_{0}}{2}\gamma B$. If we multiply this quantity by the Bohr magneton $e\hbar/ 2mc$, then we get the magnetic moment $\triangle M\left(B\right)\approx\frac{N_{0}}{2}\frac{e\hbar}{2mc}\gamma B$. Thus, $\frac{N_{0}}{2}\frac{e\hbar}{2mc}\gamma$ is a magnetic susceptibility. The effects of the magnetic field and changes in the exchange integral are similar. Then we can write $\triangle N\left(B,\triangle J\right)=\frac{N_{0}}{2}\left(1-e^{-\gamma B-\beta\triangle J}\right)=\frac{N_{0}}{2}\kappa\left(B,\triangle J\right)$. In this letter, we consider small and positive values of $\triangle J$. However, it is not difficult to generalize our reasoning to negative values of $\triangle J$. To estimate the change in the exchange integral with a change in the interatomic distance $x$, we use the Gorkov-Pitaevskii formula~\cite{Gor'kov,Herring}: $J\left(x\right)\sim\left(\frac{x}{a_{B}}\right)^{5/2}\exp\left(-\frac{2x}{a_{B}}\right)$. If the interatomic distance is greater than the Bohr radius $a_B$, then as this distance decreases (compression of the crystal lattice), the exchange integral increases. This explains the enhancement of the spin MR at those pressures when metallization and enhancement of the Lorentzian MR have not yet occurred, but due to an increase in the exchange integral, the spins have become more ordered. We are mainly interested in the region near the MR crossover. The change in the exchange integral is no longer significant near this region. Therefore, further in the article we will not use this quantity in the equations.

Let us introduce the notation: $\tau_{1}$ is the relaxation time for scattering of parallel spins and $\tau_{2}$ for scattering of antiparallel spins. Then for electrons with spin up and spin down we have the following expressions for the average scattering times
\begin{equation}
\tau_{\uparrow}^{B}=N_{0}\frac{\tau_{1}\tau_{2}}{M\tau_{1}+N\tau_{2}},  \tau_{\downarrow}^{B}=N_{0}\frac{\tau_{1}\tau_{2}}{N\tau_{1}+M\tau_{2}}
\end{equation}
With no magnetic field $N=M=N_{0}/2$. Then 
\begin{equation}
\tau_{\uparrow}^{0}=\tau_{\downarrow}^{0}=\frac{2\tau_{1}\tau_{2}}{\tau_{1}+\tau_{2}}
\end{equation}
In the presence of a magnetic field $\triangle N\neq 0$ and
\begin{align}
\tau_{\uparrow,\downarrow}^{B}=\frac{2\tau_{1}\tau_{2}}{\left(1\mp\kappa\left(B\right)\right)\tau_{1}+\left(1\pm\kappa\left(B\right)\right)\tau_{2}}
\end{align}
Using these equations, we can estimate the conductivity. In the presence of a magnetic field, the conductivity is the tensor. In the Drude approximation we have \cite{Kittel,Alisultanov}
\begin{align}
\sigma_{ixx}=\sigma_{iyy}=\frac{n_{i}e^{2}\tau_{i}}{m}\frac{1}{1+\omega_{H}^{2}\tau_{i}^{2}}\\
\sigma_{ixy}=-\sigma_{iyx}=\frac{n_{i}e^{2}\tau_{i}}{m}\frac{\omega_{H}\tau_{i}}{1+\omega_{H}^{2}\tau_{i}^{2}}
\end{align}
For total conductivity we have $\sigma=\sigma_{1}+\sigma_{2}$. Then for the total longitudinal resistance and MR we obtain
\begin{align}
\rho_{xx}=\frac{\sigma_{\uparrow xx}+\sigma_{\downarrow xx}}{\left(\sigma_{\uparrow xx}+\sigma_{\downarrow xx}\right)^{2}+\left(\sigma_{\uparrow xy}+\sigma_{\downarrow xy}\right)^{2}}
\end{align}
It is also easy to show that for $\tau_{1} \neq \tau_{2}$ and $\kappa=0$ (non-magnetic material) the Drude MR (13) is equal to zero. This is due to the fact that at $\kappa=0$ both channels are equivalent and therefore the effect of the Lorentz force on both channels is compensated by the Hall field. At the same time, for $\kappa\neq0$ MR is nonzero, but always negative. This means that in this model, the spin contribution to the MR is always greater than the Lorentzian contribution due to the rapid saturation of the latter. The Lorentzian contribution can exceed the spin one only in the presence of another mechanism that quenches the saturation of this contribution. One of these mechanisms is associated with the inclusion of a new type of carriers, holes, in the model. This is the key trick for our model.

The above calculations included only one type of carrier - electrons. However, near the metal-insulator phase transition, the electron concentration is low. On the other hand, since the phase transition temperature is about 300 K, the hole concentration is comparable with the electron density. Therefore, when studying the MR of our system, it is necessary to use the electron-hole model. In the Drude approximation, the equations of motion for this problem can be written as follows~\cite{Alekseev}
\begin{align}
\bm{\upsilon}_{\uparrow}^{e,h}=\pm\frac{e\tau_{\uparrow}}{m^{e,h}}\bm{E}\pm\frac{e\tau_{\uparrow}}{m^{e,h}c}\bm{\upsilon}_{\uparrow}^{e,h}\times\bm{B}\\
\bm{\upsilon}_{\downarrow}^{e,h}=\pm\frac{e\tau_{\downarrow}}{m^{e,h}}\bm{E}\pm\frac{e\tau_{\downarrow}}{m^{e,h}c}\bm{\upsilon}_{\downarrow}^{e,h}\times\bm{B}
\end{align}
where the indices $e,h$ refer to electrons and holes, respectively. The top sign refers to electrons and the bottom sign to holes. In our model, we assume that the relaxation times of electrons and holes for a given spin projection are the same. Moreover, further we assume that $m^{e}=m^{h}$. The solution of this system of equations gives the following results for currents
\begin{align}
j_{\uparrow x}=\sigma_{\uparrow}E_{x}\\
j_{\downarrow x}=\sigma_{\downarrow}E_{x}
\end{align}
where
\begin{equation}
\sigma_{\uparrow,\downarrow}=\frac{e^{2}n}{m}\frac{\tau_{\uparrow,\downarrow}}{1+\omega_{H}^{2}\tau_{\uparrow,\downarrow}^{2}}\left[1+\nu^{2}\omega_{H}^{2}\tau_{\uparrow,\downarrow}^{2}\right]
\end{equation}
where $n=n^{e}+n^{h}$, $\nu=\left(n^{e}-n^{h}\right)/n$, $n^{e},n ^{h}$ are the concentrations of electrons and holes. The total current is defined as the sum of the currents from both channels: $j_{x}=j_{\uparrow x}+j_{\downarrow x}=\left(\sigma_{\uparrow}+\sigma_{\downarrow}\right)E_ {x}$. This gives the following expression for the resistance
\begin{equation}
\rho_{xx}=\frac{1}{\sigma_{\uparrow}+\sigma_{\downarrow}}
\end{equation}
For $H=0$ we have $\tau_{\uparrow}=\tau_{\downarrow}=\tau^{0}$. Then
\begin{equation}
\rho_{xx}\left(H=0\right)=\frac{m}{2ne^{2}\tau^{0}}=\frac{m\left(\tau_{1}+\tau_{2}\right)}{4ne^{2}\tau_{1}\tau_{2}}
\end{equation}
Finally, for MR we obtain
\begin{equation}
MR=\frac{1}{\frac{\tau_{\uparrow}\left(1+\nu^{2}\omega_{H}^{2}\tau_{\uparrow}^{2}\right)}{1+\omega_{H}^{2}\tau_{\uparrow}^{2}}+\frac{\tau_{\downarrow}\left(1+\nu^{2}\omega_{H}^{2}\tau_{\downarrow}^{2}\right)}{1+\omega_{H}^{2}\tau_{\downarrow}^{2}}}\frac{4\tau_{1}\tau_{2}}{\tau_{1}+\tau_{2}}-1
\end{equation}
Without taking into account the Lorentz curvature of the trajectory ($\omega_{H}=0$), we obtain that MR is negative: $MR=-\kappa^{2}\left(B\right)\frac{\left(1-\alpha\right)^{2}}{\left(1+\alpha\right)^{2}}$, where $\alpha=\tau_{2}/\tau_{1}$. Figure 3 illustrates the magnetic field dependence of MR at different values of parameter $\tau_{2}$, which is expected to increase with pressure, as discussed above. The parameter $\alpha=\tau_{2}/\tau_{1}$ is assumed constant, as both $\tau_{1}$ and $\tau_{2}$ increase at similar rates. The figure shows that a change in $\tau_{2}$ results in a change in the sign of MP when $\nu$ is not equal to 1 ($\nu=1$ in the absence of holes). We set $\alpha=6$ and $\nu=0.9$, and observe that introducing only $10 \%$ of holes can cause a change in the sign of MR with increasing $\tau_{2}$. In fact, this effect also occurs even with a smaller number of holes. Although parameters $\nu$ and $\omega_{H}$ may also change with applied pressure, these changes are negligible and do not affect the qualitative behavior of the effect on MR sign change. 

\begin{figure}
    \centering
    \includegraphics[width=\columnwidth]{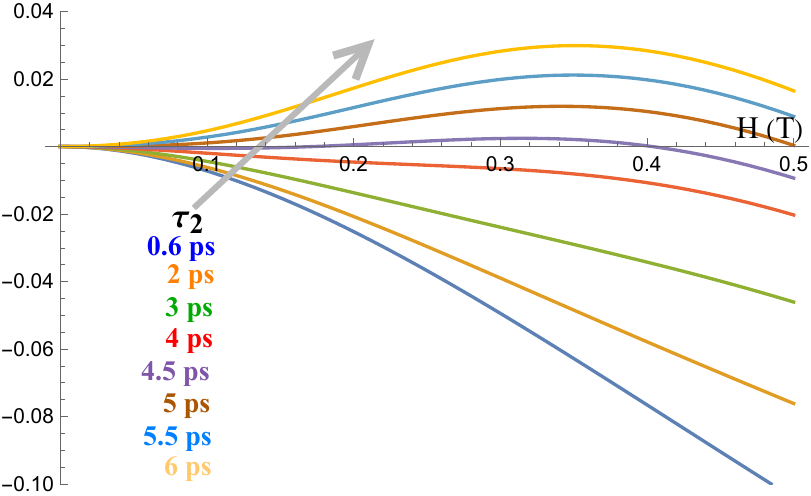}
    \caption{Magnetoresistance vs. magnetic field at different values of scattering time. The number of holes is $10 \%$.}
    \label{fig:my_label}
\end{figure}

In this letter, we reported the effect of a change in the sign of the magnetoresistance in La$_{0.8}$Ag$_{0.1}$MnO$_{3}$ under external pressure near the metal-insulator phase transition. The effect can be qualitatively explained using the two-component Drude theory, when the electrons and holes are carriers.

\section*{Acknowledgements}
Z.Z.A. acknowledges the RSF (No. 22-72-00110) and MIPT Priority-2030 Program for supporting the theoretical part of the article. A.G.G. acknowledges the Priority-2030 Program of NUST “MISiS” (No. K2-2022-022) for supporting the experimental part of the article.

\nocite{*}
\bibliography{aipsamp}

\end{document}